\documentclass[aps,twocolumn,showpacs]{revtex4}

\usepackage[T1]{fontenc}
\usepackage{graphicx}
\usepackage{color}
\usepackage{times} 

\newcommand{\Sh}{\ensuremath{\rm{Sinh}}}
\newcommand{\Ch}{\ensuremath{\rm{Cosh}}}

\begin{document}
\title{Thermal properties of AlN-based atom chips}
\author{J. Armijo, C. L. Garrido Alzar and I. Bouchoule}
\affiliation{
 Laboratoire Charles Fabry de l'Institut d'Optique, UMR 8501 du CNRS, 
91127 Palaiseau Cedex, France  }


\begin{abstract}We have studied the thermal properties of  
atom chips consisting of
 high thermal conductivity Aluminum Nitride (AlN) substrates on which 
  gold microwires are directly deposited.
We have measured the heating of wires of several widths and with different 
thermal couplings to the copper mount holding the chip.
The results are in good agreement with a theoretical model where the copper
mount is treated as a heat sink and the thermal interface resistance 
between the wire and the substrate is vanishing. We give analytical formulas
describing the different transient heating regimes and the steady state.
We identify criteria to optimize the design of a 
chip as well as the maximal
currents $I_c$ that can be fed in the wires.
For a 600~$\mu$m thick-chip glued on a copper block with Epotek H77, 
we find $I_c=16~$A for a  3~$\mu$m high, 200~$\mu$m wide-wire.
\end{abstract}


\maketitle

\section{Introduction}
A few years after the first Bose Einstein condensates (BEC) in atomic 
vapours, the will to miniaturize the set-ups has led to the production of atom 
chips, which consist in microfabricated elements, most usually wires, that are 
used to trap and manipulate cold atomic clouds. The atom chip soon became 
increasingly popular as a compact, robust and versatile device suitable for the 
production of BEC  \cite{Folman02,BecZimmermann,JacobBEC} and 
for studies on quantum matter, 
cold atom-based metrology, or quantum information \cite{FortaghRMP07}.

The major interest of atom chips is to manipulate atoms in the close  
vicinity of the field sources, so that trapping potentials with strong 
spatial variations can be obtained at very low power consumption. Small 
structures and tightly confining traps allow one to perform efficient 
evaporative cooling at high collision rates, to squeeze cold clouds 
to very anisotropic geometries so that low dimensional regimes are 
reached \cite{EstevePRL06,TrebbiaPRL06,AmerongenPRL08,Hofferberth08}, 
and make it possible to realize a great diversity of 
trapping geometries \cite{FortaghRMP07}.

Still, there are limits to the miniaturization of the structures 
and for several reasons, large currents can still be needed. First, 
as the loading stages require traps deep and wide enough to 
collect a high number of 
atoms \cite{Extavour09}, big structures running large currents 
are needed. 
These large loading structures can  be placed below the 
chip \cite{optimizedUmot04}, but other limitations 
still prevent the use of arbitrarily small structures. 
One problem is that when atoms are brought close to 
the source wire they become sensitive to the potential roughness 
created by the wire's 
imperfections \cite{EstevePRA04,Wildermuth2005}. 
Second, it is often 
desirable to avoid the  interactions between the atoms 
and the surface (Casimir-Polder force~\cite{LinPRL04,ObrechtPRL07} or
Johnson noise 
causing spin-flip 
losses~\cite{HenkelApplPhys99,Cornell-spinfliplosses}), 
which involves keeping the atoms at 
least some microns away from it. For all those reasons, the maximal 
current that can be carried by each wire is a crucial parameter 
determining the possibilities of an atom chip. 
Unless
one uses superconducting wires~\cite{nirrengarten:200405}, 
in which case   
the  maximal current is the critical current above which the metal  
becomes normal,
the maximal current 
is determined by the dissipation of the heat generated in the
 wires. In all this paper, we only consider  resistive wires, 
much simpler to achieve experimentally and most widely used.

Up to now, atom chips have mainly been realized on Silicon (Si) wafers because Si is cheap, fabrication techniques are well developed, and it is a good thermal conductor. But, as Si is semiconductor, an electrically insulating layer, generally SiO$_{2}$, needs to be placed between the wafer and the metallic wires. Unfortunately, SiO$_{2}$ is also a thermal insulator, and, in a previous study \cite{Groth04}, it was found that this layer is the main limitation to the removal of heat in Si-based atom chips. On the other hand, AlN 
is a substrate material that has been especially selected for 
being simultaneously a good electrical insulator and a good 
thermal conductor. It was first used  for high-power 
microelectronics applications \cite{Werdecker}. More and more 
groups working on atom chips are now moving to AlN substrates 
\cite{Extavour09,Lev03,Hui08,ChipGlassCell04}, because they allow 
for direct deposition of the wires on the substrate. Since 
no thermal contact resistance between the wire and the substrate is
expected, much better heat dissipation is foreseen.
 
  We have fabricated such chips and measured their thermal behavior. We 
first present the model that we have developed to understand our 
experiments and identify the different heating regimes.
We then show that this model reproduces very well 
the measured wire heating  for different wire widths, 
different thermal couplings to the copper mount and 
within the different heating regimes.
 In particular, we show that the heating 
in AlN-based atom chips is only governed by heat diffusion in the substrate, 
unlike  the Si-based atom chips. Therefore, the relevant phenomenon 
to consider for the current limitations is the long-time heating
rather than a fast heating due to a thermal resistance 
between the wire and the substrate. The 
thermal coupling to the copper mount and the possibility to quickly reach a 
stationary regime thus acquire a crucial importance.
 Finally, we draw pratical conclusions on the optimal design and the current 
limitations of AlN-based atom chips.

\section{Theoretical model}
 At any time of its operation, the temperature of a current-carrying 
wire is the result of its resistive heating  and the heat removal via different channels. Most of the heat is removed by conduction through the substrate on 
which the wire is deposited, out to the chip mount, considered as a heat reservoir. When the chip is placed  in a vacuum chamber, the only other mechanism for heat dissipation is black-body radiation of the wire which, as we show later, has a negligible effect. In air, 
as for some of the experiments presented below, air convection can
also play a role, but it is expected to be negligible, in particular since the wires in the chip we use are covered by a $6~\mu m$ thick-layer of resist of high thermal resistance. In the following, 
we thus assume that heat conduction inside the substrate is 
the only mechanism for heat removal and we compute the expected 
heating of the wire. 

\begin{figure}[htbp]
\begin{center}
\includegraphics[width=8 cm]{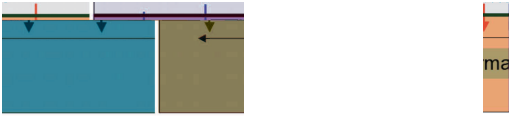}
\caption{Sketch of our model. The substrate is infinite in the $x$ direction and the Cu mount acts as a heat reservoir. The heat dissipated in the wire diffuses in the substrate in the 1Dz, 2D and 1Dx regimes successively. When the transverse spread is $l_s$, an equilibrium state is reached (see text). (color online)}
\label{fig.schema}
\end{center}
\end{figure}

\begin{table}
\begin{tabular}{|l|p{4.8cm}|l|}
\hline
$\lambda$&AlN conductivity & 128 W/(Km)\\
$D$ & $\lambda/c$~: AlN diffusion constant & 53 mm$^2$/s\\
$\sigma$ & thermal contact resistance between substrate and heat sink & 
$1.1\times 10^{-4}$ m$^2$K/W\\
$e$&substrate thickness&600~$\mu$m\\
$l_c$& $\lambda \sigma$~: contact length & 14~mm\\
$l_s$&$\sqrt{el_c}$ : stationary length& 2.9~mm\\
$t_s$ &$l_s^2/D=\sigma e c$ : stationary time & 0.16~s\\
$t_e$ & $e^2/D$ : crossover time between $2D$ and $1Dx$ regimes&6.7~ms\\
$\Phi$ & resistive power generated by the wire per unit length & 920 Wm$^{-1}$\\
\hline
\end{tabular}
\caption{Relevant parameters for the wire heating and their value 
in the case experimentally studied in this paper where the chip is glued 
onto the copper block. The value of $\Phi$ is given for a $200~\mu$m wide, 
$3~\mu$m high gold wire 
running a current of 5~A.}
\label{tab.params}
\end{table}

 In our model, depicted in Fig.~\ref{fig.schema}, 
we assume that the wire, of width 
$W$ along the $x$ direction is infinite in the $y$ direction. 
We also assume that the substrate of thickness $e$ 
is infinite in the transverse direction x.
The back surface of the substrate is supposed to be in contact 
with a heat reservoir at temperature $T_0$ with a 
thermal contact resistance 
$\sigma$ (in m$^2$K/W). 
Finally, we assume  that the energy flux per
unit area from the wire to the substrate is homogeneous over the wire width.
Using these hypothesis, we compute the temperature 
inside the substrate and in particular just below the wire. We assume that the thermal contact resistance  between the wire and the substrate is vanishing so that the substrate temperature 
at the wire position is equal to the wire temperature.

 The substrate thermal properties are described by its specific heat 
per unit volume $c$ and its thermal conductivity $\lambda$. The 
resulting diffusion constant is $D=\lambda/c$.
With the thermal contact resistance between the substrate and 
the heat reservoir, 
we can construct a length $l_c=\lambda\sigma$. However, a more relevant 
length scale is $l_s=\sqrt{el_c}$, denoted as the stationary length, 
as we explain in the following.
Thus, three different length scales govern the heat diffusion 
process : $W$, $e$ and 
$l_s$. From these lengths we derive the corresponding time scales $t_W=W^2/D$, 
$t_e=e^2/D$ and $t_s=l_s^2/D$. 
Table~\ref{tab.params} gathers the most relevant parameters and gives 
their value for the experimental case studied in this paper, in 
the most favorable situation where the chip is glued onto the copper block.

Let us first consider the 
step response of the system for a wire of constant resistivity~:
the wire and substrate are at temperature 
$T_0$ when the current is suddenly turned on, producing
a heat flow from the wire to the substrate per unit 
length and per unit time $\Phi$. 
We compute the evolution of the wire temperature, noting 
by $\Delta T$ its deviation from $T_0$. In the equations, we 
express temperature in energy unit, setting $k_B=1$. 
To give some physical insight we consider the 
situation where the characteristic times satisfy 
$t_W\ll t_e\ll t_s$, a situation usually fulfilled.
 For times $t \ll t_W$, the energy has diffused inside 
the substrate over lengths much smaller than the wire width $W$.
Then the diffusion inside the substrate is expected to be 
one-dimensional in the z direction (see Fig.~\ref{fig.schema}) and 
in this $1Dz$ regime
the wire heating can be written as
\begin{equation} 
\Delta T ^{1Dz}=2\frac{\Phi}{W}\frac{1}{\sqrt{\pi c \lambda}} \sqrt{t}
\label{eq.T1Dz}
\end{equation}
Next, for times $t_W\ll t\ll t_e$, the energy has diffused into the substrate 
over lengths much larger than the wire width but much smaller than the 
substrate thickness $e$, so that the heat diffusion is well described 
by a two-dimensional model (see Fig.~\ref{fig.schema}) and we expect the temperature to 
increase as
\begin{equation}
\Delta T^{2D}=\frac{\Phi}{2\pi\lambda}\ln
\left ( \frac{W_{\rm{eff}}^2+2Dt}{W_{\rm{eff}}^2} \right )
\label{eq.T2D}
\end{equation}
where $W_{\rm{eff}}$ is an effective width on the order of $W$.
At times $t\gg t_e$, the energy has diffused over lengths much 
larger than the substrate thickness $e$, so one expects now a 
one-dimensional model in the x direction (see Fig.~\ref{fig.schema}) 
to be an accurate description of the wire heating, which takes the form
\begin{equation}
\Delta T^{1Dx}=\Phi\frac{1}{e\sqrt{\pi c \lambda}} \sqrt{t}.
\label{eq.T1Dx}
\end{equation}  
This formula is valid as long as the energy flux per unit length 
to the reservoir $\Phi_R$
is much smaller than $\Phi$.
The heat flux to the reservoir per unit area is
$j_R=(T(x)-T_0)/\sigma$ where
$T(x)$ is the substrate temperature at position $x$. 
Using $T(0)-T_0=\Delta T$ and 
since $T(x)-T_0$ falls off to zero over a width of order 
$L=\sqrt{Dt}$, we find that $\Phi_r=\int dx j_R 
\simeq \Delta T \sqrt{Dt}/\sigma$. 
Inserting in Eq.~\ref{eq.T1Dx}, we find 
$\Phi_R\simeq t \Phi D/(el_c)=\Phi t/t_s$.
Thus Eq.~\ref{eq.T1Dx} is valid as long as $t\ll t_s$.
For $t\simeq t_s$, $\Phi_R\simeq \Phi$~: the heat flux to the reservoir 
compensates the input heat flux $\Phi$ and the system reaches
a steady state. This justifies the denotation of $t_s$ and $l_s$
as the stationary time and length respectively (see Fig.~\ref{fig.schema}).
Interestingly, $t_s$, which can be rewritten as $t_s=ce\sigma$,
does not depend on $\lambda$. 
Note also that $t_s$ is equal to  the relaxation 
time of the substrate, defined as the time constant 
of the exponential decay of the substrate temperature
after a homogeneous heating. 
This can be understood by the intuitive picture that the heat 
spreads in the substrate as long as it has not been absorbed by 
the mount.

When the stationary flow is finally established, the equilibrium temperature at the wire
can be estimated by summing up the contributions of the three successive regimes.
The contribution of the late $1Dx$ regime is found replacing
$t=t_s$ in Eq.~\ref{eq.T1Dx}: 
$T_{eq}^{1Dx} = \alpha \frac{\Phi}{\lambda} \sqrt{\frac{l_c}{e}}$, where $\alpha$ is a 
numerical prefactor. From Eq.~\ref{eq.T2D}, we find that the $2D$ regime contributes 
to an amount $T_{eq}^{2D} = \frac{\Phi}{\pi \  \lambda} ln(\frac{e}{W_{0}})$ 
where $W_{0}$ is of the order of W. 
As for the initial $1Dz$ regime, 
its contribution scales as $\frac{\Phi}{\lambda}$ 
so that 
it is possible to incorporate it in  $T_{eq}^{2D}$  by renormalizing $W_{0}$. 
Finally we get the analytic expression for the equilibrium temperature
\begin{equation}
\Delta T_{eq} = \frac{\Phi}{2 \ \lambda} \sqrt{\frac{l_c}{e}} + 
\frac{\Phi}{\pi \  \lambda} \ln\left (\frac{e}{W_0}\right )
\label{eq.Teq}
\end{equation}
where we have replaced $\alpha$ by $\frac{1}{2}$ a posteriori from the calculation that we explain below.

\begin{figure}
\label{fig.perc}
\centerline{{\includegraphics[width=7cm]{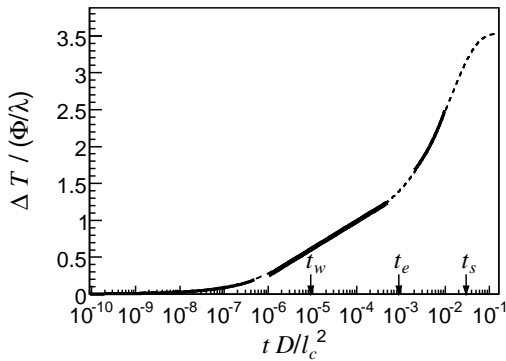}}}
\caption{Transient wire heating for $W=3\times 10^{-3} l_c$ and 
$e=3\times 10^{-2} l_c$ (dashed line). 
Temperature and time are scaled to $\Phi/\lambda$ and $t_c=l_c^2/D$
respectively. 
The solid lines correspond to fits with the functions (from 
earlier to later times) $a_1*\Delta T^{1Dz}$,
$\Delta  T^{2D}$ and $a_3+b_3\Delta T^{1Dx}$,
the fitted parameters being
$a_1=0.74$, $W_{\rm{eff}}=0.17 W$,  
 $a_3=1.0$ and $b_3=0.79$.}
\label{fig.reponseechelon}
\end{figure}

In order to get quantitative predictions and to investigate the crossovers between 
the different diffusion regimes, it is useful to perform numerical calculations.
For this purpose we use standard Fourier analysis. 
For an input heat flow at the top of the substrate that varies
as $j_{k,\omega}e^{ikx}e^{i\omega t}$, we find the temperature at $x,z=0$
\begin{equation}
\Delta T_{k,\omega}=j_{k,\omega} \frac{\Sh(Ke)/l_c+K\Ch(Ke)}{\Ch(Ke)/l_c+K\Sh(Ke)}
\frac{1}{K\lambda},
\label{eq.Tkomega}
\end{equation} 
where $K=\sqrt{k^2+ic\lambda\omega}$. 
We make the assumption that the heat flow j(x) is homogeneous over the wire size. The equilibrium temperature is then obtained 
setting $\omega$ to 0 in Eq.~\ref{eq.Tkomega}.
In the case $l_c \gg e \gg W$, the wavevectors much smaller than $1/e$ give 
a lorentzian contribution, whose integral yields the 
prefactor $\alpha = 1/2$ for the first term in rhs of Eq.~\ref{eq.Teq}.

Most importantly, we use Eq.~\ref{eq.Tkomega} to compute the transient heating of the wire. Fig.~\ref{fig.reponseechelon} shows the result of the calculation for an input energy flux $\Phi$ which is a step function in time, for 
well separated length scales.
Fits in the three different temporal domains are in good agreement
with the analytical laws derived above.  
 We observe that the initial $1Dz$ behavior  fails noticeably already 
for $t\simeq t_W/10$ and that $W_{\rm{eff}}$ in Eq.~\ref{eq.T2D}
is noticeably smaller than $W$ ($W_{\rm{eff}}\simeq 0.17(3) W$).
 On the other hand, the  transition from the $2D$ to the $1Dx$ regime
and the convergence to the equilibrium state occur 
as expected for $t\simeq t_e$ and $t\simeq t_s$ respectively. 

 The above calculations assume that the energy flow $\Phi$ is 
constant in time, however, as the wire heats up, its resistivity,
and consequently $\Phi$ as well, also 
increase.
To take this effect into 
account numerically, we compute the 
wire temperature as 
 \begin{equation}
 T(t)=T_0 + \int_0 ^t \!  R(t-t') \Phi(t') \, dt'
 \label{eq.calcT}
 \end{equation}
where the impulse response function $R(t')$ is determined using Eq.~\ref{eq.Tkomega}. 
To compute $\Phi(t')$ at a given $t'$, we 
assume a homogeneous wire temperature, so 
that $\Phi(t') =I^2 \rho(T(t') )/ h W$. This 
approximation is justified a posteriori 
since the computed wire temperature inhomogeneity 
never exceeds 10\% for the parameters explored in this paper.
The dependence of the gold resistivity with temperature is approximated by 
 \begin{equation}
\rho=\rho_0(1 + \alpha \Delta T)
\label{eq.alpha}
\end{equation}
where 
$\alpha = 0.0038~\rm{K}^{-1}$ is obtained from a linear fit of the 
reported values of $\rho$
between 200~K and 500~K~\cite{CRC08}.

Our last remark about the model concerns the assumption that the 
input energy flux $j(x)$ is homogeneous over the wire width. This 
is expected to be true if there is no spatial redistribution of the 
energy released by Joule's effect inside the wire, but might be 
inaccurate if the wire height is not small compared to its width or 
if the wire thermal diffusion constant is very large compared to 
that of the substrate. In the extreme case where the wire temperature 
is uniform, simple calculations can be done, to provide an upper bound on
the error. More precisely, for $e\gg W$ and for times much larger 
than $t_W$, we find\footnote{
On distances 
much smaller than $\sqrt{Dt}$, where $t$ is the typical timescale 
considered, the temperature distribution obeys the two-dimensional 
Laplace equation $\Delta T=0$. Using the complex variable $x+iz$
and a Schwartz-Christoffel conformal transformation, we find that 
$T \propto {\cal R}\rm{e}(\rm{ArcCosh}(x+iz))$. The energy flux from the wire 
is $j(x)=-\lambda\left . \partial T/\partial z\right )_{z=0}.$
} $j(x)\simeq 2\Phi/(\pi W\sqrt{1-(2x/W)^2})$.
Using Eq.~\ref{eq.Tkomega}, we then find that the equilibrium 
temperature is lower than in the case of a uniform $j(x)$ by an amount 
smaller than 5\% for all the parameters explored in this article. Thus, 
we expect that the error in our predictions caused by our assumption on 
$j(x)$ cannot be larger than 5\%. Alternatively, the 
small wire temperature inhomogeneity predicted by 
our model shows that both boundary conditions give similar results.

\section{Comparison with measurements}
 We now come to the experimental measurements and their comparison 
with the model. The  chip is based on an AlN substrate of thickness
600(50)~$\mu$m and of size 25~mm$\times$35~mm. Gold wires of height 
3~$\mu$m are deposited by evaporation on the substrate, on top of a 
30~nm thick-titanium adhesion layer. We measure the wire heating  
after a constant current is turned on by monitoring the wire resistance 
and using Eq.~\ref{eq.alpha}.
The  current supplies we use allow a current rise time of 
about 10~$\mu$s. The wire resistance is deduced from 
the voltage drop across it.
The contribution $R_c$ of the
connecting wires and the contact resistances has to be 
subtracted from the measured resistance. To compute $R_c$, 
we measure the circuit resistance at a current low enough 
to produce negligible heating and subtract 
the  contribution of the microwire, computed using 
the nominal wire dimensions and the gold resistivity value  
 at 300~K.
 We perform heating measurements on two different wires : 
a 200~$\mu$m wide, 20 mm long-wire and a 7~$\mu$m wide, 3~mm long-wire.

 To investigate the heating on long time scales  
we use the 200~$\mu$m wide-wire, in which we run a current of about 5~A.
Three main cases are considered. First,
the chip is laid on paper to thermally insulate it from the copper block.
Second, it is maintained on the copper block, either in air or in vacuum.
Finally, the chip is glued on the copper block with Epotek H77.

\begin{figure}
\label{fig.200micronswire}
\centerline{{\includegraphics[width=8cm]{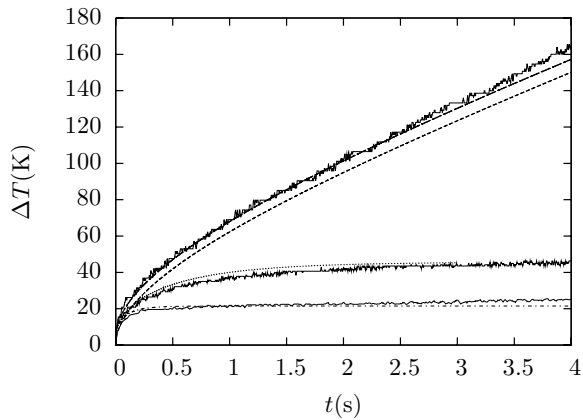}}}
\caption{Wire heating at long times for a 200~$\mu$m
wide-wire running a current of 5~A. Solid lines give the experimental results 
for, from top to bottom, a chip laid on paper, a chip 
maintained on a copper block with a pressure of $2 \times 10^3$~Pa
and  a chip glued on the copper block. The other lines 
are theoretical predictions (see text).}
\end{figure}

In the case where the substrate is laid on paper, 
we compare the data to the theoretical model, assuming 
that no heat escapes the substrate ($l_c\rightarrow \infty$).
As shown  in Fig.~\ref{fig.200micronswire}, we find agreement 
within a few percent for times lower than 2~s, 
provided the AlN thermal conductivity is set to $\lambda=128$~W/(Km).
This value is close 
to the reported values for AlN \cite{Werdecker}
although somewhat smaller. 
The discrepancy  may compensate for imprecisions on the other parameters 
(AlN heat capacity, 
gold resistivity, wire size and substrate thickness) which are fixed
to their nominal values. 
In all the following, we use this fitted value  
of $\lambda$.
With our substrate $t_e \simeq 7$~ms, so that for the times considered in Fig.~\ref{fig.200micronswire}, one expects the heating process to be well explained by the $1Dx$ 
model of Eq.~\ref{eq.T1Dx}. Using this simple model and including 
the dependence of gold resistivity with temperature, we obtain 
the short-dashed  curve shown in Fig.~\ref{fig.200micronswire}. 
It agrees within 5\% with the 
more complete calculation. The discrepancy, which is about a constant 
offset, is due to the early $2D$ regime. At times longer than  2.5~s, 
the data show an excess heating compared to the model. 
We attribute this effect to the finite size of the substrate~: after $t=2$~s
the heat has diffused over a typical distance $\sqrt{Dt}=1.1 $~cm,
larger than the distance from the wire to the substrate edge (1.0 cm).

 In the case where the substrate is maintained on copper, the relevant new
parameter is the  thermal contact resistance $\sigma$ 
between the substrate and the copper block.
We have measured $\sigma$ in the following way. 
First, we heat the substrate to a temperature 
$T_1 \simeq 500$~K
by running current in a wire and
letting the substrate thermalize for a few seconds, while keeping it isolated from the 
copper with some paper.
Then, we suddenly remove the paper so that the substrate is in thermal 
contact with the copper block, 
with an applied pressure of $2\times 10^3$~Pa. We monitor the 
thermal relaxation of the substrate by recording the resistance of 
one of the chip wires in which a small current of $10$ mA is constantly 
flowing. 
 A fit to the function 
$T=T_0+(T_1-T_0)exp(-t/\tau)$ yields the 
relaxation time $\tau =  \sigma ce= 0.82 ~s$, which  
gives the thermal contact resistance 
$\sigma=5.77\times 10^{-4}~$m$^2$K/W. 
To our knowledge no measurement of $\sigma$
was previoulsy reported for such a small contact pressure at a Cu-AlN interface.
 A much better coupling could be obtained for pressures
higher than $10^5$~Pa~\cite{Dirke08}, but such high pressures do not seem 
realistic in the context of atom chips, because usually a large free 
surface on the chip is required.
The value of $\sigma$ also depends on the surface quality, so 
we need to mention that the substrate we use has a backside 
roughness amplitude of about 100~nm. 
Importing in the model the measured value of $\sigma$, we 
compute the heating of the wire after a 5~A current-step, with no 
adjustable parameters. The result, shown as a dotted 
curve in Fig.~\ref{fig.200micronswire} agrees with the measurement 
to better than 10\%, which is a good validation for our model.

In real experiments, unless the chip is itself one of the walls of the 
vacuum chamber \cite{ChipGlassCell04}, the contact surface between the 
chip and the heat sink is in vacuum.
We have thus also studied the case of a chip maintained 
on a copper block in vacuum, at a pressure lower than $10^{-1}$~mbar.
We find that the thermal contact resistance $\sigma_{\rm vac}$ between the substrate and the copper is much larger than in air, meaning that 
the thermal coupling in room conditions is actually caused by the air 
present in the voids 
between the substrate and the copper block.
To measure $\sigma_{\rm vac}$, we monitor the cooling of the substrate 
with the same protocol as in air, with the difference that,
since the cooling time is larger than the thermalization time 
of the substrate (a few seconds), thermal insulation of the 
substrate in the initial heating stage is not needed.
The measured relaxation time is $\tau_{\rm vac}=26~$s, corresponding to
$\sigma_{\rm vac}= 1.8\times 10^{-2}$m$^2$K/W.
We also measure the relaxation time in vaccuum for a chip laid on paper
and find $82~$s, a value several times larger than $\tau_{\rm vac}$. 
This confirms that 
conduction to the copper is the main cooling mechanism for a chip 
held on copper, 
even in vacuum where the thermal 
resistance $\sigma_{\rm vac}$ is high, and that black-body radiation is negligible.
In practice, since the stationary time $t_s$ is identical to the relaxation 
time $\tau$, the coupling to the copper block has negligible effect on
the wire heating for times 
much smaller than $\tau$. In particular, within the time of our 
measurement (4s), we observed a heating equal within 5\% to the one observed
 when the chip is laid on paper.

In order to mitigate the wire heating, especially if operating in vacuum, 
the previous studies show that it is  highly desirable
to fill the voids between the chip and its mount. That is why, in the last situation, we have 
glued the chip onto the copper block with a thin layer of Epotek H77, 
a glue of high heat conductivity (according to the specifications, 
$\lambda_{H77}=0.66~$W/(mK)) containing grains whose maximal radius is 
specified to 20 $\mu m$. The experimental 
value of the steady state temperature (see Fig.~\ref{fig.200micronswire}) 
is obtained, within our model, for a  thermal contact resistance 
$\sigma=1.1\times  10^{-4}$~m$^2$K/W, which corresponds to a glue 
layer of 73~$\mu$m, assuming a homogeneous glue layer and neglecting 
interface resistances. 
Importing 
this value into the model, we compute the expected wire heating, 
shown in Fig.~\ref{fig.200micronswire} as a dashed-dotted line. It matches  the time 
evolution very well.

\begin{figure}
\centerline{{\includegraphics[width=8cm]{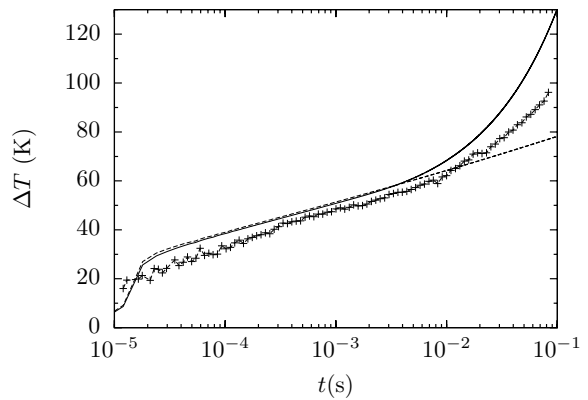}}}
\caption{
Wire heating at long times for a 7~$\mu$m
wide-wire running a current of 1.7~A.
 Lines are predictions of the 
full model (solid) and the bare $2D$ model of Eq.~\ref{eq.T2D} 
(dashed). The crosses are experimental data.}
\label{fig.2D}
\end{figure}

We now turn to the measurements of the heating at early times.  
With a 200~$\mu$m wide-wire, the early $2D$ heating regime is barely 
visible, because the condition $W \ll e$ is not fulfilled. 
We thus use the 7~$\mu$m wide-wire. 
The measurements were taken for a substrate laid on paper and are 
shown on Fig.~\ref{fig.2D}. The prediction of the 
model with no free parameters, taking into account the  10~$\mu$s 
current rising time, is shown as solid line. 
For times between 100~$\mu$s and 5~ms, we clearly 
observe a logarithmic increase of the temperature, as expected from the 
$2D$ model, and the calculation agrees within 10\% with the measurement.
The dashed line is the result of the
bare $2D$ model of Eq.~\ref{eq.T2D}, including the temperature 
dependence of gold resistivity. $W_{\rm{eff}}$ is adjusted to $0.15\,W$, 
so that the bare $2D$ model and the more complete model agree within 1\% at 
times smaller than $5$~ms.
At larger times, the model predicts a stronger
heating due to the onset of the $1Dx$ regime, 
which is also visible in the experimental data.
For times larger than 50~ms, the model overestimates the 
temperature increase rate.
This is very likely due to the finite length of the wire~: indeed
 at $t = 100$~ms, the heat has spread in 
the substrate over a typical distance $\sqrt{Dt}=2.3~$mm that is no 
longer negligible compared to the wire length $L= 3$~mm. 
At longer times, the spread of the heat becomes two-dimensional 
in the plane of the substrate, which is more efficient for  heat removal. 

Because it is experimentally relevant, it is worth investigating briefly the effect of the finite wire length $L$ on the stationary temperature. 
As long as  $l_s\ll L$, 
the equilibrium temperature is barely affected by the wire finite 
length. On the contrary, in the limit $l_s\gg L$, the model of an infinite wire fails. 
In this case, for times 
$t\gg L^2/D$, the wire temperature increases as 
$\Delta T^{2Dxy}(t)\simeq(\Phi L/4\pi e\lambda) \ln (Dt/L^2)$ until the equilibrium 
temperature is reached. Equating the input power 
with the heat transferred to the heat sink, we find that 
the equilibrium temperature is about
$T_{eq}^{2Dxy}=T_{eq}^{1Dx} \times (L/\pi l_s) \ln (l_s/L)$, a 
value much smaller than $T_{eq}^{1Dx}$ (see section 2). 
Quantitative predictions are obtained 
using a two-dimensional version of Eq.~\ref{eq.Tkomega}, 
setting $\omega$ to 0 and 
integrating over wavevectors in the x and y directions.
In the experimental case considered in this paper, 
for a chip glued to the copper block and 
for our 3~mm long-wire (for which $l_s=0.96 L$), 
$T_{eq}$ is decreased by about 25\% compared 
to its value for an infinite wire.

The data from Fig.~\ref{fig.2D} can additionally be used to place an upper 
limit on the thermal contact resistance $\sigma_W$ between the wire and 
the AlN substrate. In presence of a finite $\sigma_W$ , the wire temperature 
increases by $ \sigma_W \Phi/W $ on a time scale
$\tau_W = \sigma_W C_{Au} h$, typically smaller than 1 $\mu s$, 
where $C_{Au}$ is the heat capacity of gold. This fast heating 
has been observed in Si-based atom chips \cite{Groth04}. Our data 
however are compatible with a vanishing $\sigma_W$, since they agree 
with our model which assumes $\sigma_W =0$. More precisely, within the 
precision of our model and measurements, we can confidently say that 
the excess  heating due to a finite $\sigma_W$ cannot be more than 
10~K in the first 100~$\mu$s, which yields the upper 
limit $\sigma_W=2.3 \times 10^{-8}$Km$^2$/W, 
corresponding to a conductance of $4.3 \times 10^7$~W/(Km$^2$).  
This contact resistance is a factor 6.6 smaller than the values 
reported in ~\cite{Groth04}.

\section{Practical consequences}
  In order to minimize the wire temperature, 
it is of course desirable to minimize the thermal contact 
resistance $\sigma$ between the substrate and the reservoir. 
However, once a technology is chosen, so once $\sigma$ and $l_c$ are fixed, 
it is possible to use our model to optimize the chip design and to compute 
the maximal currents that can  flow in each wire.

\begin{figure}
\begin{center}
\includegraphics[width=8cm]{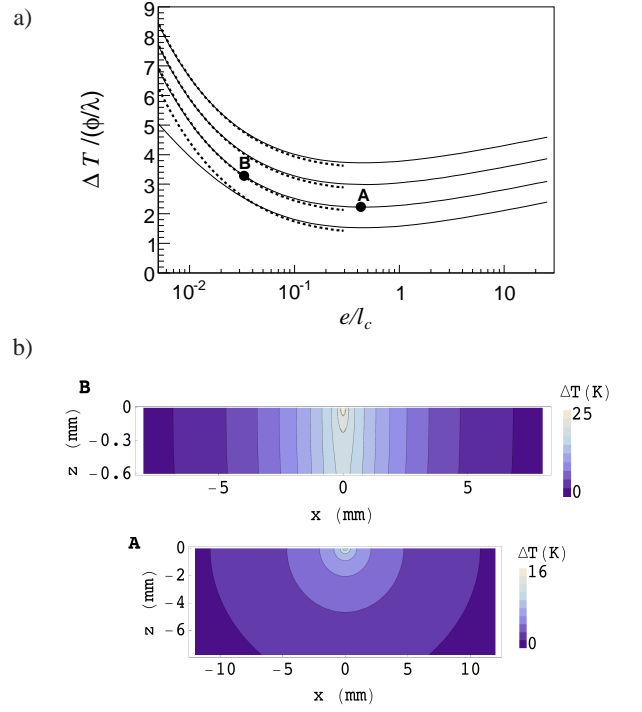}
\end{center}
\caption{Optimum substrate thickness.
a) Equilibrium wire temperature versus substrate thickness for
wire widths (from top to bottom) $W/l_c =10^{-4}$, $10^{-3}$, $1.1\times 10^{-2}$, and 0.1. 
The dashed lines correspond to Eq.~\ref{eq.Teq} with $W_{\rm{eff}}=0.61\,W$. 
Heating is minimum for a substrate thickness of 0.3~$l_c$.
b) Stationary temperature distribution
for the parameters 
of our chip glued on copper, for a substrate of thickness 
$e$=0.6 mm (B) and for a substrate of optimal thickness $e^*= 7.8$~mm (A).
The wire, 200~$\mu$m wide, runs a current of 5~A and its
temperature increase is 24.3~K in case B and 15.8~K in case A.}
\label{fig.Teq}
\end{figure}

 As we have shown from our measurements, in good operating conditions, a 
stationary state is expected to be reached within a few 100 ms. This 
means that typical experiments on atoms will fall into the stationary regime; for example, an evaporation to BEC typically takes a few seconds in our set-up. 
Therefore the stationary temperature $\Delta T_{eq}$ is the relevant parameter to consider.
In Fig.~\ref{fig.Teq}.a we plot $\Delta T_{eq}$
for different wire widths and substrate thicknesses.
The scalings used are justified by dimensional analysis~: since heat conduction inside the substrate is governed by linear 
equations, the equilibrium temperature is given by
\begin{equation}
\Delta T_{eq}=\frac{\Phi}{\lambda} f({W}/{l_c},{e}/{l_c})
\label{eq.Teqnorm}
\end{equation}
Note that the function $f$ does not depend on $c$, which only enters into account for the time scales of the transient regimes. 
One sees in Fig.~\ref{fig.Teq}.a that $\Delta T_{eq}$ is very well described by Eq.~\ref{eq.Teq}, in its domain of validity ($W\ll e\ll l_c$.
Here $W_0=0.61W$ has been obtained by fitting the calculation 
for  $W=10^{-4}l_c$ and $5\times 10^{-3}<e/l_c<0.3$.
The failure of Eq.~\ref{eq.Teq} for the lowest curve of 
Fig.~\ref{fig.Teq}.a at small $e/l_c$ is due to the fact that for these parameters
$W$ is no longer  small compare to $l_s$. Then, the $1Dx$ regime 
barely exists and only the initial $1Dz$ regime is present.
Roughly speaking, in the case $e \ll l_c$,  the temperature 
distribution has a 1D character, as can be seen in  Fig.~\ref{fig.Teq}.b(B)
which shows the temperature distribution inside the substrate for 
the parameters of Table 1. In this regime,
it is favorable to increase $e$, to "deconfine" the energy spread.
On the other hand, for $e\gg l_c$, the $1Dx$ regime of diffusion no 
longer exists~:
the steady state is realized directly after the $2D$ regime when
the energy reaches the substrate's lower surface. $\Delta T_{eq}$ is 
then of the order of $\ln(e/W)\Phi/\pi\lambda$, increasing with the substrate 
thickness (Fig.~\ref{fig.Teq}.a).
These arguments predict an optimal substrate thickness 
of the order of $l_c$. This is confirmed by the calculations 
(Fig.~\ref{fig.Teq}.a), which 
show that $\Delta T_{eq}$ reaches a minimum for the optimal substrate 
thickness $e^*\simeq0.3 l_c$, which is about independent of $W$. In this 
optimal case, the three length scales $l_s$, $l_c$ and $e$ are almost equal.
In fact, $e^*$ is the smallest value of $e$ 
that permits the suppression  of the $1Dx$ regime and therefore
the temperature distribution inside the substrate has a 
two-dimensional character (Fig.~\ref{fig.Teq}.b(A)).
 Note that  the wire heating minimum is rather
broad (Fig.~\ref{fig.Teq}.a)~: even with a substrate thickness as low as $e^*/10$, the 
heating is increased by only about 50\%.

\begin{figure}
\centerline{\includegraphics[width=7cm]{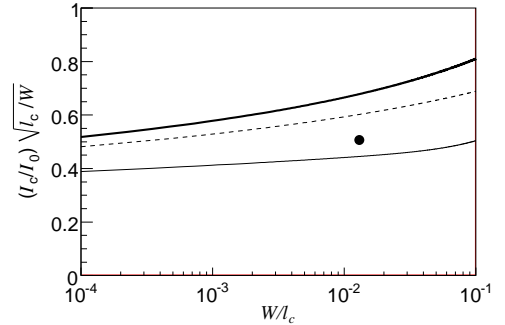}}
\caption{Critical current divided by $\sqrt{W/l_c}$ and normalized 
to $I_0=\lambda\sqrt{\sigma h/\rho_0\alpha}$, versus  wire width for 
substrate thicknesses $e/l_c=0.5$ (thick line, optimal case), 
10 (dashed line) and 0.01 (thin line).
The point represents the case studied in this paper~: 
a 200~$\mu$m wide, 3~$\mu$m high-wire on an AlN substrate 
of width $e=600~ \mu$m, for the parameters of Table 1.
 In this case, $I_c =$ 16 A.
}
\label{fig.inst}
\end{figure}

The second important quantity we can derive from  our model is the maximal
current that can flow in the wires.
Using Eq.~\ref{eq.alpha} and 
$\Phi=\rho I^2/Wh$, we find that
Eq.~\ref{eq.Teqnorm} is a self-consistent equation, whose solution 
is $\Delta T_{eq}=\rho_0 I^2/(Wh)f/(1-f l_c I^2/W I_0^2)$
where $I_0=\lambda\sqrt{\sigma h/\rho_0 \alpha}$. 
When $I$ approaches the critical current $I_c=I_0\sqrt{W/(l_cf)}$
a divergence 
occurs, signature of an instability. 
In Fig.~\ref{fig.inst}, we show 
the critical current $I_c$, where the dependence on $\sqrt{W}$ 
has been removed by dividing by  $\sqrt{W/l_c}$. 
Note that this maximal current corresponds to 
an established stationary state.
For situations where currents are only required for a time shorter
than $t_s$, higher currents can be used.

\section{Conclusion}
 
To conclude, we have presented the first study on the thermal 
properties of AlN-based atom chips. The main result is that, as 
expected, the thermal behavior is more favorable than with Si chips.
More precisely, our measurements are compatible with the absence of 
thermal contact resistance between the wire and the substrate.
The heating of the wire is  entirely explained by 
heat diffusion inside the 
substrate and its absorption by the heat reservoir holding the chip.
We have developped a model that accounts well for our
experimental data
for wires of different widths, for different couplings to the copper block and 
within the different heating regimes. 
The thermal coupling to 
the heat sink holding the chip is a crucial parameter~: in particular
we have shown that, when 
operating in vacuum, it is recommended to glue the chip to
the heat sink.
Finally, the model is used to derive pratical learnings~: the optimum 
substrate thickness is computed as well as the maximum  current 
that can be run into the wires in the stationary regime.

 We have treated a simplified case, but, in practical situations, it 
is important to 
take other effects into account.
First, as we already noticed,  the finite length of the wire 
may reduce significantly the heating.
The vicinity of wires may also affect 
the thermal behavior~:  several current-carrying wires 
separated by 
distances on the order or smaller than the stationary length $l_s$, 
will experience a stronger heating.
In the temporal domain, the accumulation of 
heat over several experimental cycles will increase the heating 
if the dead time when currents are off 
is not long compared to $t_s$.
Finally,  heat conduction through the copper block and its cooling 
may affect the heating on long timescales. Active cooling of the copper 
block may be envisioned.

 Different technologies  may improve the thermal properties of atom chips.
An expensive solution would be
to use diamond substrates because diamond,  
although electrically insulating, has a much higher
thermal conductivity than AlN. 
Another key point is 
the realization of a better thermal contact between the substrate and 
the heat sink. If the backside of the chip is in air, using a thermal grease
seems highly desirable. Working in vacuum, a process leading to a thinner glue 
layer may be developped
and/or a different glue may be used. One could also consider 
soldering, the difficulty being to avoid damaging  the chip.

The Atom Optics group of Laboratoire Charles-Fabry is a member 
of the IFRAF Institute. This work was supportedd by the ANR grant 
ANR-08-BLAN-0165-03 and by EU under the grant IP-CT-015714.
 

\end{document}